\documentclass[useAMS,usenatbib]{mn2e}
\usepackage{epsfig}          
\usepackage{graphicx,color}        
\usepackage{amssymb} 
\usepackage{color}           
\usepackage{url}
\usepackage{amsmath}
\usepackage{setspace}
\usepackage{rotating}
\usepackage{float}
\usepackage{textcomp}
\usepackage[english]{babel}
\usepackage{graphicx}
\usepackage{adjustbox}
\usepackage[mediumspace,mediumqspace,squaren]{SIunits}
\hbadness=\maxdimen
\vbadness=\maxdimen

\usepackage{natbib}
\usepackage{tabularx}

\title[ Precessing Radio Jets in a DPAGN]
  {A Candidate Dual AGN in a Double-peaked Emission-line Galaxy with Precessing Radio Jets}
\author[Rubinur K. et al.]{Rubinur, K.$^{1,2}$\thanks{E-mail: rubinur@iiap.res.in}, M. Das$^1$,  P. Kharb$^3$, M. Honey$^{1,2}$ \\
  $^1$Indian Institute of Astrophysics, Bangalore 560034, India\\
  $^2$Pondicherry University, R. Venkataraman Nagar, Kalapet, 605014 Pondicherry, India\\
  $^3$National Centre for Radio Astrophysics - Tata Institute of Fundamental Research, \\
Pune University Campus, Post Bag 3, Ganeshkhind Pune 411007, India}

\def\LaTeX{L\kern-.36em\raise.3ex\hbox{a}\kern-.15em
    T\kern-.1667em\lower.7ex\hbox{E}\kern-.125emX}

\begin{document}

\label{firstpage}

\maketitle

\begin{abstract}
We present high resolution radio continuum observations with the 
Karl G. Jansky Very Large Array at 6, 8.5, 11.5 and 15 GHz of 
the double-peaked emission-line galaxy 2MASXJ12032061+1319316.
The radio emission has a prominent S-shaped morphology with highly symmetric radio jets
that extend over a distance of $\sim1.5^{\prime\prime}$ (1.74~kpc) on either side of the
core of size $\sim0.1^{\prime\prime}$(116~pc). The radio jets have a helical structure resembling
the precessing jets in the galaxy NGC~326 which has confirmed dual active galactic nuclei (AGN). 
The nuclear bulge velocity dispersion gives an upper limit of $(1.56\pm$0.26$)\times$10$^8$~M$_{\odot}$
for the total mass of nuclear black hole(s). We present a simple model of precessing jets 
in 2MASXJ1203 and find that the precession timescale is around 10$^5$ years: this matches 
the source lifetime estimate via spectral aging. We find that the expected 
super massive black hole (SMBH) separation corresponding to this timescale is 0.02 pc.
We used the double peaked emission lines in 2MASXJ1203 to determine an orbital speed 
for a dual AGN system and the associated jet precession timescale, which turns out to be
more than the Hubble time, making it unfeasible. We conclude that 
the S-shaped radio jets are due to jet precession caused either by a binary/dual SMBH system,
a single SMBH with a tilted accretion disk or a dual AGN system 
where a close pass of the secondary SMBH in the past has given rise to jet precession.

\end{abstract}


\section{Introduction} \label{section1}
According to the ${\Lambda}\mathrm{CDM}$ Universe, massive galaxies form
through the mergers of smaller galaxies \citep{Springel.etal.2005}. 
During mergers, the super massive black holes (SMBHs) of the individual galaxies
sink to the center of the merger remnant \citep{begelman.etal.1980} and form a SMBH pair. 
Simulations show that mergers are accompanied by massive gas inflows \citep{Hopkins.etal.2009}.
This can result in accretion of gas onto the SMBHs resulting in dual active galactic nuclei (DAGN) 
with SMBH separations of 0.1-10 kpc or binary AGN with separations $<100$~pc \citep{spolaor.2015}.
The gas accretion may also give rise to radio jets that launch along 
the direction of black hole spin \citep{Rees.etal.1982}. 
The binary SMBHs will eventually coalesce producing huge amount of gravitational waves \citep{Thorne.etal.1976}. 
Theoretical studies predict that DAGN should be common,
but the number of resolved DAGN is low  mainly because the required resolution is very hard to achieve.
To date, there are only $\sim$23 resolved DAGN \citep{Deane.etal.2014, Mullersanchez.etal.2015}.

The resolved DAGN are detected through high resolution imaging at radio, X-ray, 
optical or Near infrared (NIR) wavelengths. There is only one confirmed binary AGN 
which is the radio galaxy 0402+379. In this case, the Very Long Baseline Array (VLBA) 
has resolved the binary at 7~pc separation \citep{Rodriguez.etal.2006}. On kpc scales, 
the few convincing examples are LBQS 0103-2753 \citep{Junkkarinen.etal.2001}, 
NGC~6240 \citep{Komossa.etal.2003}, 3C~75 \citep{Hudson.etal.2006}, Mrk\,463 \citep{Bianchi.etal.2008},
Mrk\,739 \citep{Koss.etal.2012}. More examples are in \citet{Fu.etal.2011,Liu.etal.2013,Comerford.etal.2015}.
In optical and NIR, the detection of two cores in imaging is not sufficient because one of the cores can
be a starburst nucleus or a stellar bulge. Therefore, spectroscopic observations of the cores are 
required along with imaging \citep{McGurk.etal.2015}.

There are a few indirect signatures of DAGN/binary AGN: (1)~Periodicity in optical variability; 
(2)~Double-peaked AGN (DPAGN) emission lines in optical nuclear spectra; (3)~S- or X-shaped radio morphology. 
The initial detections were serendipitous and the earliest DAGN were detected from 
their optical variability \citep[e.g. OJ287][]{sillanpaa.etal.1988,Lehto.etal.1996}. 
Recently a close supermassive black-hole binary in quasar PG 1302-102 has been discovered
from optical periodicity \citep{Graham.etal.2015}. The other example is PSO J334.2028+01.4075 \citep{Liu.etal.2015}.

DPAGN are one of the signature of DAGN \citep{Zhou.etal.2004}. 
Since the advent of high resolution spectroscopic surveys (e.g. SDSS)\footnote{www.sdss.org}, 
large samples of candidate DAGN have been identified from DPAGN spectra \citep{liu.etal.2010}.
But DPAGN can also be due to rotating disks \citep{Greene.etal.2005,muller.etal.2011,Kharb.etal.2015}, 
jet-ISM interaction or outflows \citep{Whittle.etal.2005,Rosario.etal.2010}. 
Therefore, in order to confirm dual AGN in DPAGN galaxies, one has to carry out high resolution imaging
at different wavelengths. At present, almost 30\% confirmed DAGNs are from DPAGN samples
\citep[e.g.,][]{Fu.etal.2011, McGurk.etal.2011}. 
\citet{Tingay.etal.2011} observed 11 DPAGN using VLBI but did not find any double radio cores.
They concluded that DPAGN may not be a good indicator of dual AGN. However, 
\citet{McGurk.etal.2015} have pointed out that the contribution of
DPAGN sample in identifying DAGN is significant. In recent studies, additional observations 
- such as long-slit spectroscopy,  have been used for detecting DAGN in high resolution 
radio imaging observations \citep{Comerford.etal.2012,Mullersanchez.etal.2015}.

The presence of S- or X- shaped radio jets was suggested to be connected with binary SMBHs 
by \citet{begelman.etal.1980}. Merger of two SMBHs can give rise to inversion symmetry 
in radio jets (S, Z and X shaped sources) \citep{Rottmann.etal.2001,Komossa.etal.2006,Gergely.etal.2009,Mezcua.etal.2011}.
Jet axis reorientation in the presence of a companion galaxy can be the reason for 
peculiar radio morphologies \citep{Wirth.etal.1982}. 
The interaction between the DAGN can result in precessing jets that appear as 
X- or S-shaped radio sources \citep{Merritt.etal.2002,Zier.etal.2002}. 
\citet{Ekers.etal.1978} observed NGC\,326 which is a Z-shaped radio source and explained 
the morphology by jet precession. NGC~326 is now a confirmed dual AGN system \citep{murgia.etal.2001,Hodges.2012}.
Another example is the micro-quasar SS433. This is a famous  S-shaped radio source and it is a 
binary system \citep{Blundell.etal.2004}. \citet{Gopal.etal.2003} have suggested that Z- or S- shaped sources
are SMBHs that are close to coalescence and are associated with gravitational radiation,
whereas X-shaped radio sources are merged systems and have radiated away gravitational waves. 
But warping of the accretion disks \citep{Pringle.etal.1996} or back flowing gas \citep{Leahy.etal.1984} 
can also produce some signature of X- or S- shaped galaxies.

We find evidence of S-shaped radio jets in the double-peaked AGN  of 2MASXJ12032061+1319316 (2MASXJ1203 hereafter).
In Section~\ref{section2} we describe the target galaxy. 
The observations and data reduction are described in Sections~\ref{section3} and \ref{section4}. 
In Section~\ref{section5a} we describe the results of our 6, 8.5, 11.5 and 15~GHz Expanded Very Large Array (EVLA) observations.
2D composition of galaxy image is described in Section~\ref{section5b}. 
Further estimation of SMBH mass and other parameters are discussed in Sections~\ref{section5c}. In Section~\ref{section5d},
we describe the spectral index map. We also present a simple model of the precessing jets and a rough
estimate of the precession timescale in Section~\ref{section5e}.
In Section~\ref{section5f}, we estimate the minimum magnetic field and lifetime of the source from the equipartition theorem.
In Section~\ref{section6} we discuss the origin of the S-shaped morphology and DPAGN emission from 2MASXJ1203. 
We summarize our conclusions in Section~\ref{section7}. 
Throughout this paper we assume a value of $\Omega_{m}=~0.27$, and $H_{0}=~73.0$~km~s$^{-1}$~Mpc$^{-1}$.
The spectral index, $\alpha$, is defined such that the flux density at frequency $\nu$ is S$_\alpha\propto~\nu^\alpha$.

\section{THE TARGET GALAXY} \label{section2}
\begin{figure}
\centering
\includegraphics[bb=80 350 550 730,scale=0.35,width=\columnwidth]{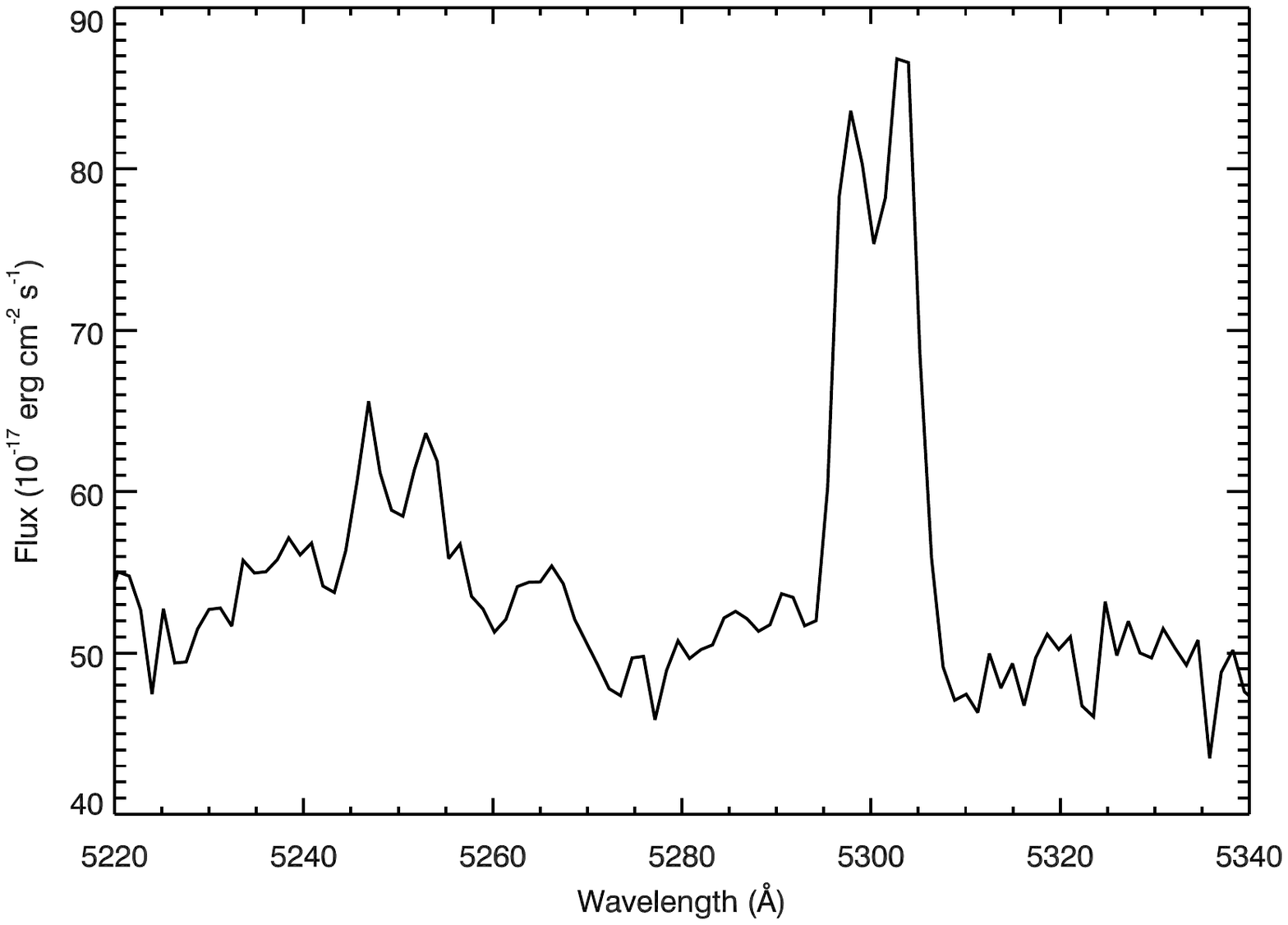}
\caption{\small The  double peaked emission lines in the SDSS optical spectrum from the nuclear region of SDSSJ120320.7+131931. 
The lines are [OIII, $\lambda$5007$\angstrom$] and [OIII $\lambda$4959$\angstrom$] of the [O III] doublet.
There is no broadening which suggests the absence of AGN outflows.}
\label{fig1}
\end{figure}

The galaxy 2MASXJ1203 is part of a larger ongoing study of DPAGN at 6~GHz using the EVLA (Project ID: VLA/15A-068)
(Rubinur K. et al. 2016B in preparation). Our preliminary data analysis at 6~GHz revealed that
the nuclear emission had an interesting S-shaped morphology and two hotspots.
On analysing additional archival data at higher frequencies (8.5 and 11.5~GHz),
we clearly resolved the compact core and found S-shaped radio jets.
2MASXJ1203 has been studied earlier at optical wavelengths in 
DPAGN surveys \citep{Fu.etal.2012,Ge.etal.2012,Wang.etal.2009}. 
The galaxy has a moderate redshift (z~=~0.058), a compact core and extended disk (Table~\ref{table1}). 
It may be an S0 or spiral galaxy \citep{Fathi.etal.2010} but its HyperLeda \footnote{http://leda.univ-lyon1.fr/}
classification has a large error \citep{Makarov.etal.2014}.
In Section~\ref{section5b}, we discuss our own 2D galaxy image decomposition using GALFIT. 
The optical spectrum shows double peaks in [O III], [O II], H${\alpha}$, [N II] emission lines (Figure~\ref{fig1}).
The underlying stellar velocity dispersion is 189.9 km~sec$^{-1}$ (see Section~\ref{section4}).
The nucleus shows radio emission in both  NRAO VLA Sky Survey (NVSS)\footnote{http://www.cv.nrao.edu/nvss/}
and Faint Images of the Radio Sky at Twenty-cm (FIRST)\footnote{http://www.cv.nrao.edu/first/} images at 1.4~GHz.
The NVSS peak flux density is 0.108~Jy at 1.4~GHz. The nucleus has a Seyfert~2 classification and 
is radio-loud with a radio loudness parameter of 156 \citep{Fu.etal.2012}. 
The Rossi All Sky Survey (RASS) \footnote{https://heasarc.gsfc.nasa.gov/docs/xte/XTE.html} 
covered the source region during one of its survey scans for about 440~s;
but 2MASXJ1203 is not detected in the RASS faint source catalogue
which has a flux limit of $F_{x}\leq1\times10^{-13}$ ~erg~cm$^{-2}$~s$^{-1}$. 
The galaxy was observed by GALEX Survey for 107s in the near ultraviolet (NUV) band; 
the NUV flux density is $9.60\times10^{-6}$~Jy. NASA/IPAC Extragalactic Database (NED)\footnote{http://ned.ipac.caltech.edu/}
shows that there are two galaxies within the velocity range from 17015~km~s$^{-1}$ to 18015~km~s$^{-1}$.
However, the galaxy does not show any signatures of interaction such as tidal tails in its optical image and
hence may not be undergoing a major merger. A minor merger, however, cannot be ruled out.

\section{OBSERVATION AND ARCHIVAL DATA } \label{section3}
We observed 2MASXJ1203 on 20 July 2015 (Project ID: 15A-068)~(Table~\ref{table2}) in the C-band at 6~GHz using the A 
configuration (resolution~0.33$^{\prime\prime}$) for 5 minutes along with the flux calibrator, 
J0542+4951 for 10 minutes. The nearest phase calibrator, J1239+0730 was observed for 50 seconds. 
The observations were done with a 1792~MHz wide baseband centered at 5.935~GHz with fourteen spectral windows, 
each of which have 64 channels with frequency resolution of 2~MHz. We also reduced the 8.5 GHz and
11.5~GHz archival data of 2MASXJ1203 (Project ID:VLA/13B-020). The observations were carried out in the X-band,
with two frequency centers  at 8.5 GHz and 11.5 GHz in the A configuration with eight spectral windows. 
We have carried out EVLA 15~GHz observations of 2MASXJ1203 in the B array configuration (resolution~0.42$^{\prime\prime}$)
(Project ID: 16A-144) on 29th May 2016. 2MASX1203 was observed for 16 minutes along with the flux calibrator 3C~286 and 
phase calibrator J1224+2122. 3C~286 was observed for ~11 minutes and J1224+2122 was observed for 8 minutes. 

\begin{figure}
\centering
\includegraphics[bb=70 360 550 740,scale=0.40,width=\columnwidth]{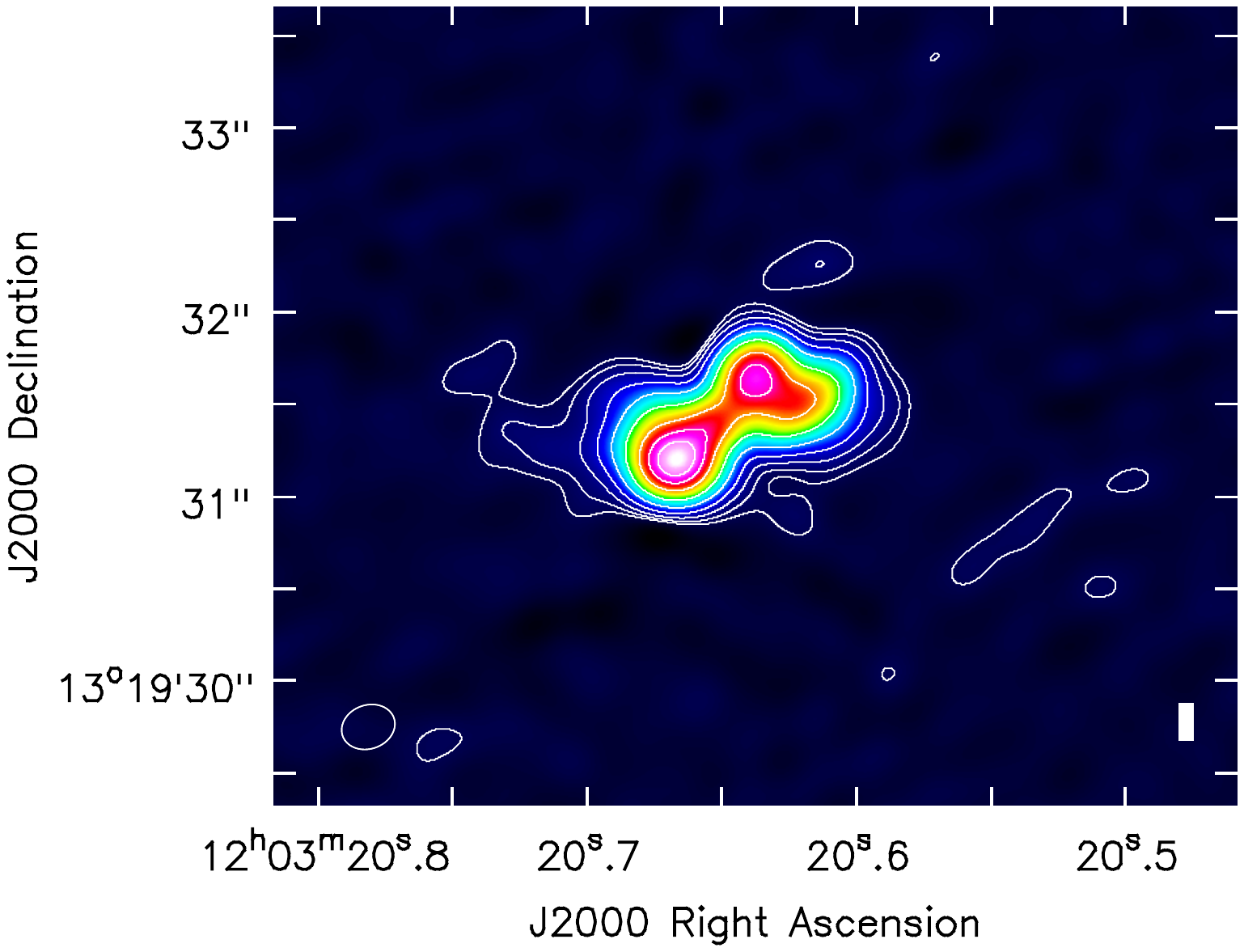}
\caption{\small The uniform weighted 6 GHz EVLA image of 2MASXJ12032061+131931.
The beam size is $0.29\arcsec \times 0.24\arcsec$. The rms noise in the image is $\sim20~\mu$Jy. 
The contour levels correspond to 0.60, 1.25, 2.5, 5, 10, 20, 40, 60, 80 \% of peak flux density at 11.90~mJy.  }
\label{fig2}
\end{figure}

\section{ Data Reduction} \label{section4}
We have used the Common Astronomy Software Applications (CASA) \citep{McMullin2007} and
Astronomical Image Processing System (AIPS) packages for data reduction. 
The bad data was flagged to obtain good solutions. We used PLOTMS to identify the bad data and the task
FLAGDATA to flag it. We have used the task GAINCAL to obtain the calibration solutions and applied it using the task APPLYCAL.
After a satisfactory calibration we imaged the source using the task CLEAN. We made both natural
and uniform weighted images using the Briggs robust parameter 0.5 and $-0.5$ respectively. 
The natural weighted images recovered most of the flux density but gave a poor spatial resolution;
hence we used it to derive the total extent of the radio jets. The uniform weighted maps gave a better spatial resolution
that enabled us to distinguish between the core and the jets. For the 6, 8.5 and 15~GHz maps, 
we obtained good images after one round of phase self-calibration. For 11.5~GHz,
we obtained a better image after two runs of phase self-calibration.
We have used IMFIT and IMSTAT tasks to get the core size, peak flux density and noise (Table~\ref{table3}).

We have used the 8.5~GHz and 11.5~GHz images to generate the spectral index map (Figure~\ref{fig8}).
We made a new 11.5~GHz image with same restoring beam as the 8.5~GHz map by constraining the beam size 
in the task CLEAN. We used the task COMB in AIPS and task IMMATH in CASA to obtain the spectral index map and 
spectral index error map. The spectral index image was created after blanking flux density values below $3\sigma$,
at both frequencies. We followed the similar procedure to make spectral index map with 6~GHz and 15~GHz images (Figure~\ref{fig9}).

We have used the SDSS DR~12 I-band image for checking the morphology of 2MASXJ1203
using {\sc GALFIT} \citep{peng2010,peng2002}.
The SDSS images were first converted back into count units using a {\sc IDL} program. 
We created the Point Spread Function (PSF) and masked the near by bright sources in the frame to get an accurate fitting. 
The residual images are produced in {\sc GALFIT} by subtracting the image made by the convolution of the model
with PSF  from the original galactic image. 

 We have used the pPXF (Penalized Pixel-Fitting stellar kinematics extraction) code \citep{Cappellari2004}
to calculate the velocity dispersion of the underlying stellar population in the nucelar region,
which uses the Gauss-Hermite parametrization technique \citep{Gerhard1993}. 
We have used the SDSS DR12 spectrum for this task (Figure~\ref{fig1}). 
The code first masks the emission lines in the spectrum and then fits a model to the absorption lines 
using a combination of single stellar population templates of different ages. It starts with initial guess
values of the galaxy systemic velocity and stellar velocity dispersion ($\sigma$). 
The best fit model gives the stellar velocity dispersion value $\sigma$~=~189.9~km~s$^{-1}$.

\section{ Results} \label{section5}
\begin{figure}[]
\centering
\includegraphics[bb=80 350 550 730,scale=0.50]{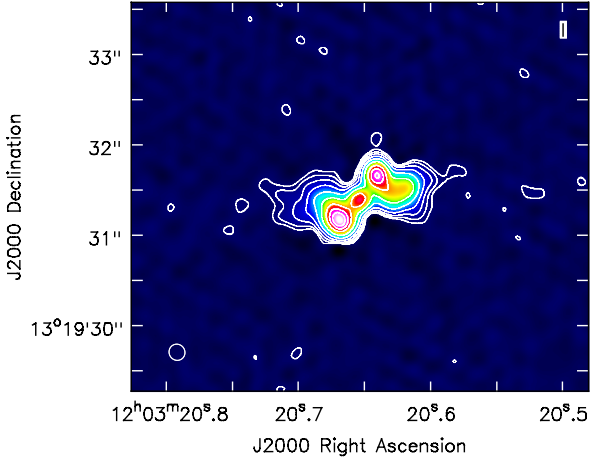}
\caption{\small The uniform weighted 8.5 GHz EVLA image of 2MASXJ12032061+131931.
The beam size is $0.18\arcsec \times 0.17\arcsec$. The contour 
levels correspond to 0.60, 1.25, 2.5, 5, 10, 20, 40, 60, 80 \% of peak flux density value at 6.86~mJy.}
\label{fig3}
\end{figure}


\subsection{Radio Images} \label{section5a}

\begin{figure}
\centering
\includegraphics[bb=100 360 450 770,scale=0.53]{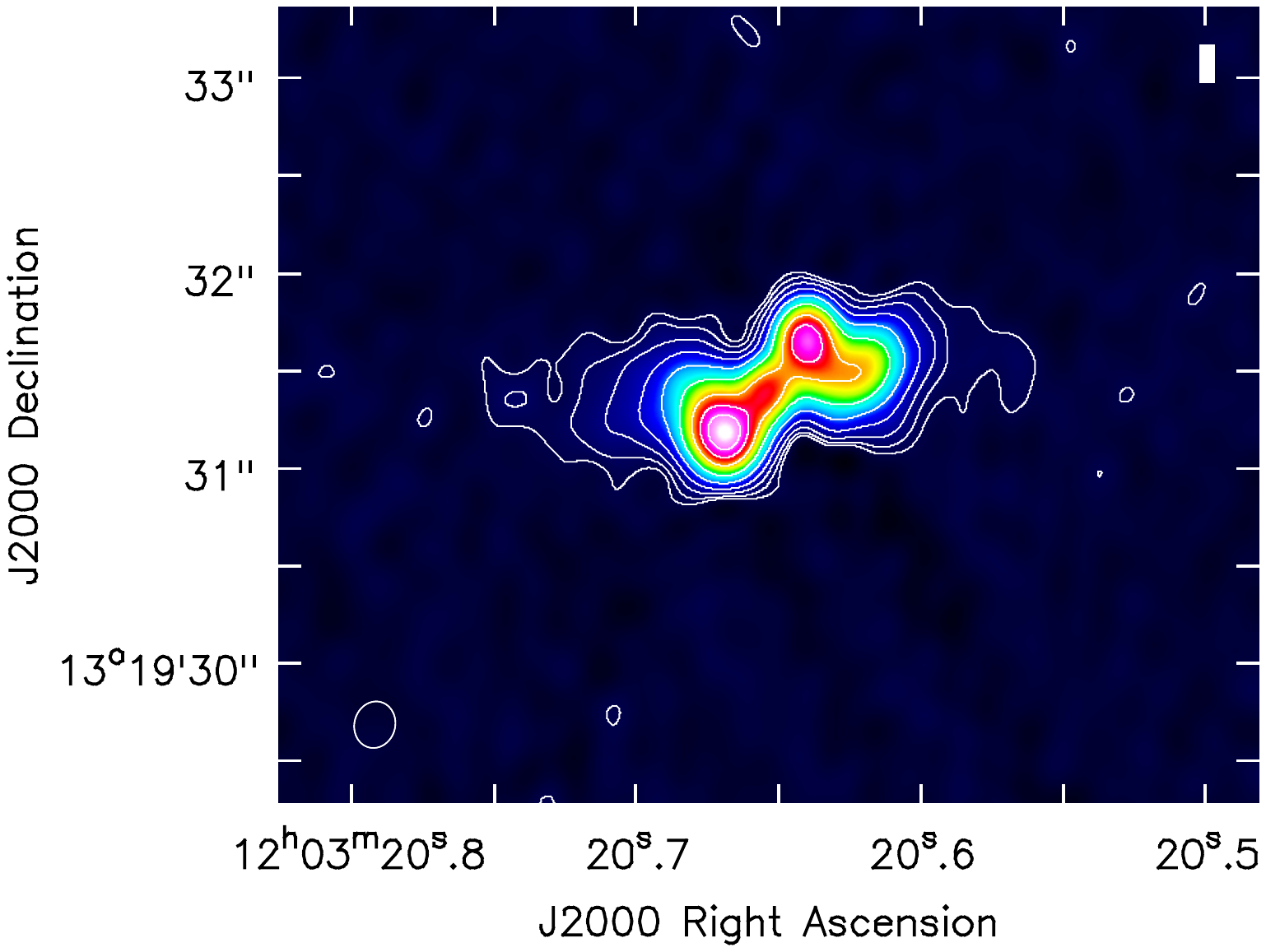}
\caption{\small The naturally weighted 8.5 GHz EVLA image of 2MASXJ12032061+131931.
The beam size is $0.24\arcsec \times 0.21\arcsec$. The contour 
levels correspond to 0.60, 1.25, 2.5, 5, 10, 20, 40, 60, 80\% of peak flux density at 8.20~mJy.}
\label{fig4}
\end{figure}

\begin{figure}
\centering
\includegraphics[bb=55 350 550 770,scale=0.55,]{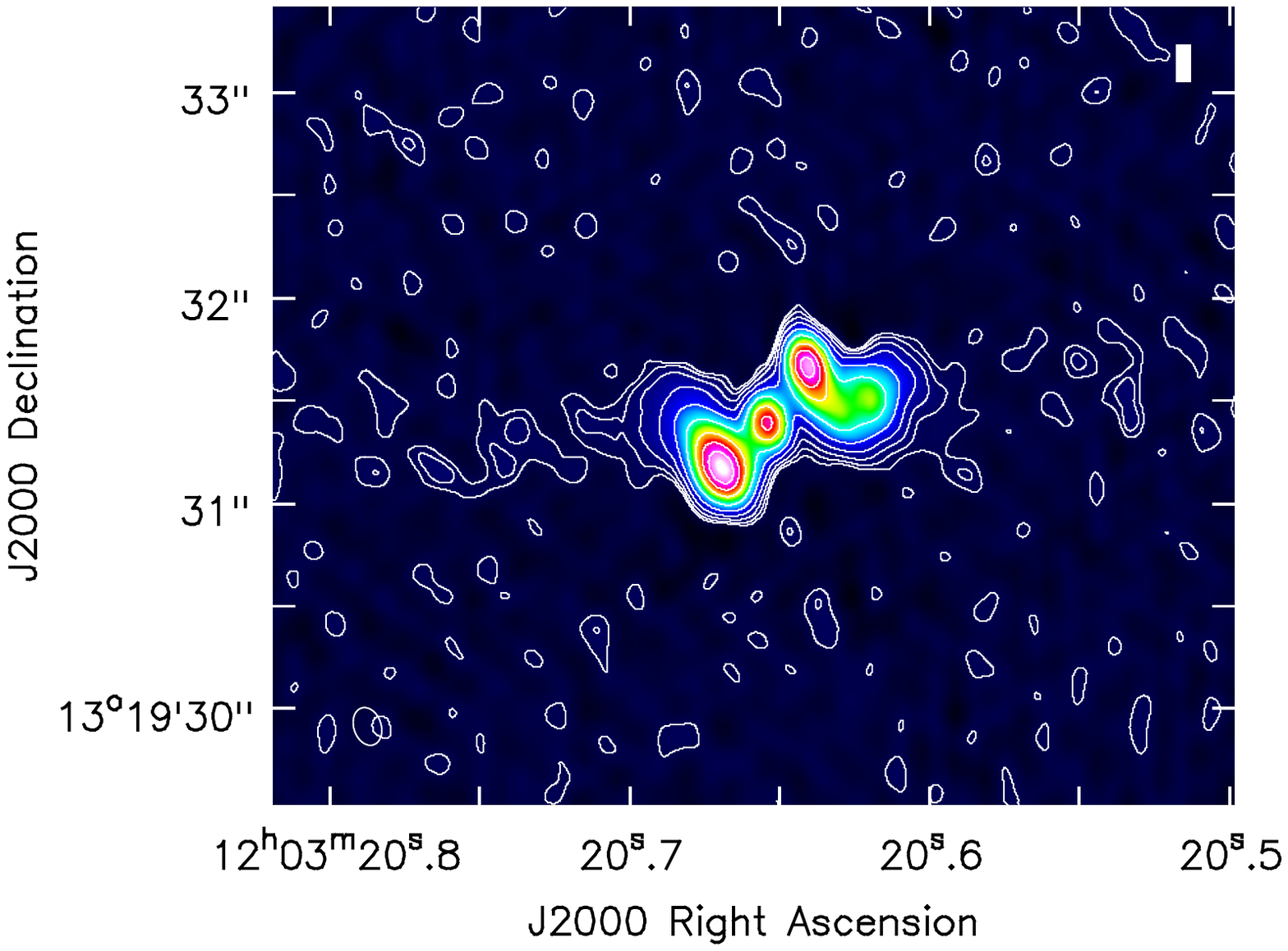}
\caption{\small The uniform weighted radio contour map 2MASXJ12032061+131931 at 11.5 GHz.
The beam size is $0.19\arcsec \times 0.13\arcsec$. The contour levels correspond to
0.60, 1.25, 2.5, 5, 10, 20, 40, 60, 80\%   of peak flux density of value 4.90 mJy.}
\label{fig5}
\end{figure}

\begin{figure}
\centering
\includegraphics[bb=50 375 550 730,scale=0.60,width=\columnwidth]{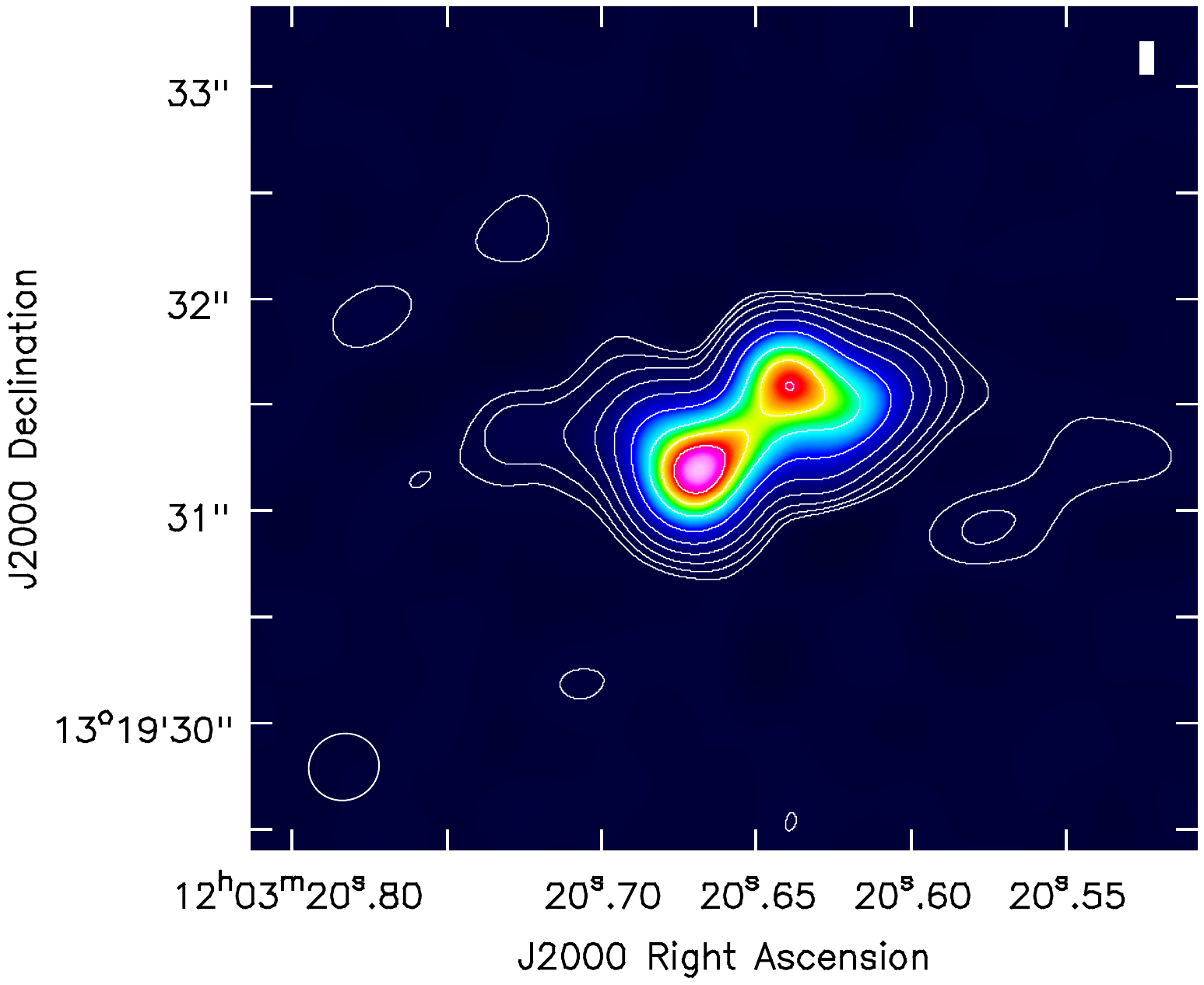}
\caption{\small The naturally weighted radio contour map 2MASXJ12032061+131931 at 15 GHz.
The beam size is $0.33\arcsec \times 0.31\arcsec$. The contour levels correspond to
0.60, 1.25, 2.5, 5, 10, 20, 40, 60, 80\% of peak flux density value of 7.30 mJy.}
\label{fig6}
\end{figure}

The radio images at 6, 8.5, 11.5 and 15~GHz are shown in Figures~\ref{fig2},~\ref{fig3},~\ref{fig4},~\ref{fig5} and \ref{fig6}. 
The uniform weighted images gave the optimum spatial resolution and flux densities 
but we used naturally weighted map at 8.5~GHz to examine the S-shaped structure  (Figure~\ref{fig4}). 
The 6 and 15~GHz radio images have two distinct radio lobes and hotspots on either side of the nucleus (Figure~\ref{fig4} and \ref{fig6}).
The hotspot south of the nucleus has a larger flux density of 11.9~mJy (see Table~\ref{table3})
suggesting that it is the closer jet and is curved towards the east. 
The increased brightness could be indicative of doppler boosting. The 8.5~GHz (Figure~\ref{fig3}) and 11.5~GHz (Figure~\ref{fig5})
images reveal a core lying between the radio lobes. The core-jet structure is completely resolved at these frequencies
and the deconvolved core has a size of $\sim0.1^{\prime\prime}$ or 116~pc (Table~\ref{table3}).
However, at 8.5~GHz  the naturally weighted image (Figure~\ref{fig4}) shows the full extent of the helical S-shaped jet 
structure out to a radius of approximately  $\sim$1.5$^{\prime\prime}$ or 1.74~kpc.
Interestingly, both the high and low resolution maps show distinctive curved radio jets,
thus producing the S-shaped morphology for the radio source. The helical structure of the 
jets is very clear since the hotspots lie along the NW-SE direction, but the 
extended radio emission lies along the NE-SW direction.

\subsection{2D decomposition of galaxy image} \label{section5b}
\begin{figure*}
\includegraphics[bb=10 420 580 780,clip,scale=.76]{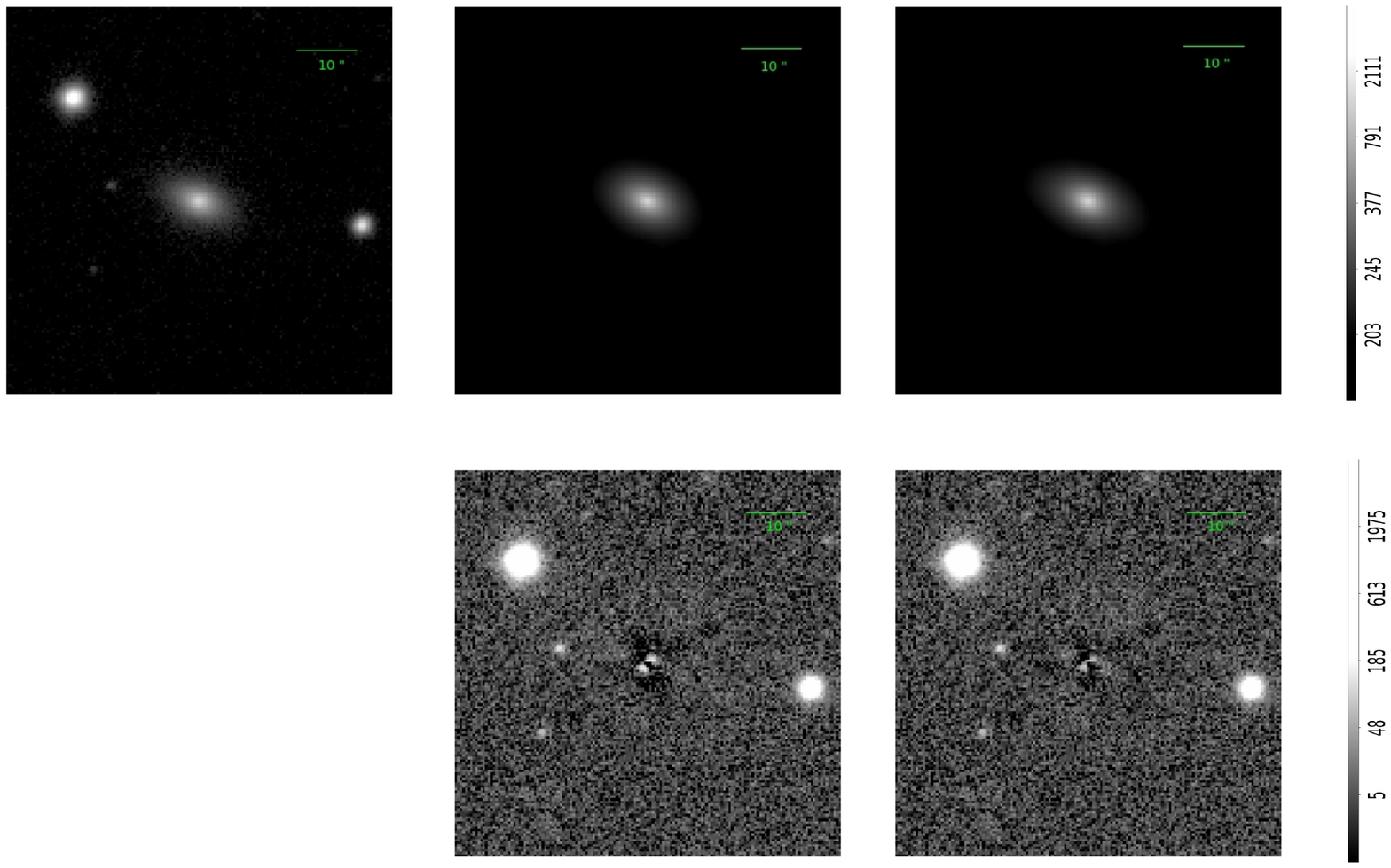}
 \caption{\small  The optical image of the host galaxy of 2MASXJ1203 is fitted 
 with Sersic and exponential profiles. Starting from the top left panel 
 and moving from left to right, is the original galaxy image, the model 
 with a single Sersic profile, and the model with a Sersic plus an exponential profile. 
 The second row presents the residual images from the models presented immediately above. 
 The color bars at the extreme right apply to all images in the respective rows. In all images,
 North is to the top and East towards the left.}
\label{fig7}
 \end{figure*}

Dual AGN  are rare in spirals, and S-shaped radio sources are usually associated with major mergers 
of elliptical galaxies (e.g. NGC~326). Hence, it is important to see if 2MASXJ1203 is an elliptical 
or spiral galaxy. To determine the morphological class of 2MASXJ1203, 
we used  GALFIT to first fit a single Sersic profile (which is used for elliptical galaxies) to the SDSS I band image.
The fit is shown in Figure~\ref{fig7}. The fit appears good and has a $\chi^2$ value of 1.097.
However according to \citet{Fathi.etal.2010} 2MASXJ1203 could be an S0 type galaxy.
Therefore, we tried with two components simultaneously - a Sersic profile for a bulge and an exponential profile
for a disk. While the two component fit results in an identical $\chi^2$ value,
the residual image looks better compared to the single component fit. 
GALFIT therefore favours the disky morphology for the host galaxy. 
However, the exposure time for the SDSS image is very short. 
A deeper image is required to confirm the morphological class of 2MASXJ1203.

%
%
%
%


\subsection{SMBH mass, Eddington ratio and Star Formation Rate} \label{section5c}
We have used the velocity dispersion of the underlying stellar component to calculate the total mass
of the central bulge which is also an upper limit to the mass of the dual SMBHs. 
Using the nuclear stellar velocity dispersion $\sigma$ that we derived (Section~\ref{section4}) from the SDSS~DR12 spectrum 
(Figure~\ref{fig1}) and the M-${\sigma}_\star$ relation in \citet{McConnell.etal.2013}, 
we obtained an upper limit of $M_{BH}=(1.56\pm0.26)\times10^8M_\odot$ for the SMBHs. Furthermore, 
we used an approximate model of a  binary  system  with  the components $M_A$ and $M_B$ in 
circular orbits about a common center of mass to obtain a very rough approximation for the individual SMBH masses.
There are several assumptions in this calculation - (i)~we have assumed that there has been a major merger
and the bulges have relaxed and hence the M-${\sigma}$ relation is valid, (ii)~the BHs are rotating in a Keplerian disk
with a separation of 0.1-2~kpc and the [O~III] peaks trace their relative velocity, 
(iii)~both SMBHs are accreting at a similar Eddington rate. According  to  Kepler's  law, 
for a  binary  system the orbital velocity ratio of the two components is given by \citet{Wang.etal.2009}

\begin{eqnarray} \nonumber
V_A/V_B=M_B/M_A=\Delta\lambda_A/\Delta\lambda_B =2.57/2.37=1.084.
\end{eqnarray}
Here $V_A$ and $V_B$ are the Doppler redshift and blueshift velocities
of the [O III] line. $\Delta\lambda_A$ and $\Delta\lambda_B$ are the Doppler redshift and blueshift
of the [O III] line in units of $\angstrom$ which have values of 2.57 and 2.37 respectively \citep{Wang.etal.2009}. 
Using the total mass to be $(1.56\pm0.26)\times10^8M_\odot$ the individual SMBHs
are $\sim0.82\times10^8M_\odot$ and $\sim0.74\times10^8M_\odot$. 
In general, however, a constant separation between the SMBHs will not perturb the accretion disks
and produce AGN activity; also the orbits are more likely to be elliptical than circular.
We note that this calculation only gives a rough estimate of the individual SMBH masses 
since there is no clear signature of a major merger in the optical image, as mentioned in Section~\ref{section2}.
The separation is also not confirmed as we will discuss later in Section~\ref{section6b}

We have obtained the blueshifted ($16.6\times10^{-16}$ ~erg~cm$^{-2}$~s$^{-1}$) and
redshifted ($19.1\times10^{-16}$ ~erg~cm$^{-2}$~s$^{-1}$) 
narrow-line [O~III]$\lambda$5008 fluxes from \citet{Ge.etal.2012}.
The total flux density is $35.7\times10^{-16}$ ~erg~cm$^{-2}$~s$^{-1}$. 
The [O~III] luminosity turns out to be $2.79\times10^{40}$ ~erg~s$^{-1}$. 
We have estimated the bolometric luminosity ($L_{bol}$) using the relation from \citet{Heckman.etal.2004}:
$L_{bol} /L_{[O~III]} \approx$ 3500 and it's value is  $L_{bol} = 9.79\times10^{43}$ ~erg~s$^{-1}$. 
For a BH mass $(1.56\pm0.26)\times10^8M_\odot$ , the Eddington luminosity ($\equiv1.25\times10^{38} M_{BH} /M_\odot$) 
is $\approx1.87\times10^{46}$ ~erg~s$^{-1}$. The Eddington ratio in 2MASXJ1203 ($â¡ L_{bol} /L_{Edd})$ is $\sim0.0052$
which is typical for Seyfert galaxies \citep[e.g.,][]{Ho.etal.2008}.

We have derived the star formation rate (SFR) for the nuclear region in 2MASXJ1203 
using the UV flux and $H_\alpha$ line emission. We have assumed a Salpeter initial mass function (IMF)
\citep{Salpeter.etal.1955} and stellar mass limits of 0.1 to 100 $M_\odot$ \citep{Kennicutt.etal.1998}. 
The UV SFR is $\sim0.105~M_{\odot}$yr$^{-1}$ for the GALEX NUV flux of our target which is 9.60~$\mu$Jy. 
We have obtained the $H_\alpha$ flux from SDSS nuclear spectrum of 2MASXJ1203 and 
it has a value of $25.3\times10^{-16}$~erg~cm$^{-2}$~s$^{-1}$.
This yields a nuclear SFR of $\sim0.156~M_{\odot}$~yr$^{-1}$.
The full $H_\alpha$ luminosity is from both star formation as well as from AGN activity.
In order to obtain the SFR, we need to subtract the AGN contribution from the $H_\alpha$ line luminosity. 
The contribution from AGN activity depends on selection criteria \citep{Fujita2003}.
We have used \citet{Tresse1996} to get the AGN contribution in low redshift galaxies which is 8\% to 17\%.
We have used these limits to subtract the AGN contribution and calculate the SFR using only stellar emission. 
The corrected SFR range is $0.130-0.138$~M$_{\odot}$~yr$^{-1}$. 
The GALEX UV flux is from the entire galaxy and the SDSS H$_\alpha$ emission is from 
the central region ($\sim$3$^{\prime\prime}$ diameter SDSS fibre). 
The fact that both the UV flux and $H_\alpha$ line fluxes give similar SFRs indicates that the star formation
 is confined to the central region of the galaxy. Similar SFRs have been derived
from infrared data in other Seyfert galaxies \citep[e.g.,][]{kharb.etal.2016}.

We also checked the optical variability of 2MASXJ1203 using 
the Catalina Real-time Transient Survey (CRTS)\footnote{http://nesssi.cacr.caltech.edu/DataRelease/} 
\citep{Drake.2009}. CRTS is a large sky, optical survey which publishes the light curves of sources
within minutes of their observations. Thus the flux variation of transient sources 
with time can be obtained from CRTS data. However, for 2MASXJ1203 no clear periodicity is evident from its CRTS data.

\subsection{Spectral index} \label{section5d}
\begin{figure}
\centering
\includegraphics[bb=30 320 550 750,clip,scale=0.460]{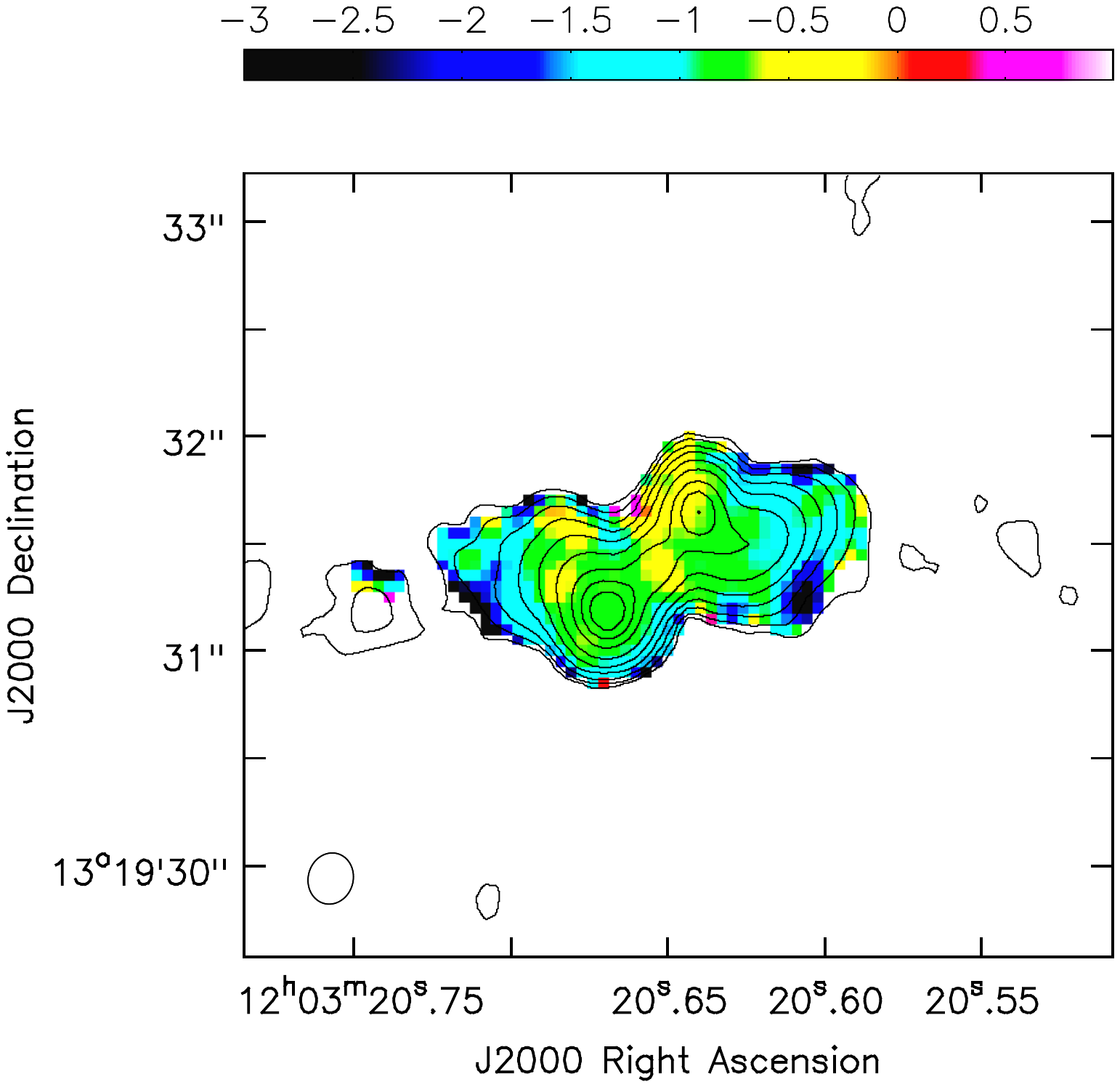}
\caption{\small The 8.5-11.5 GHz spectral index image in colour, 
superimposed by 8.5 GHz radio contours with levels corresponding to
60, 1.25, 2.5, 5, 10, 20, 40, 60, 80\%  of peak flux density value 
at 8.2~mJy.}
\label{fig8}
\end{figure}

\begin{figure}
\centering
\includegraphics[bb=30 320 550 740,scale=0.46]{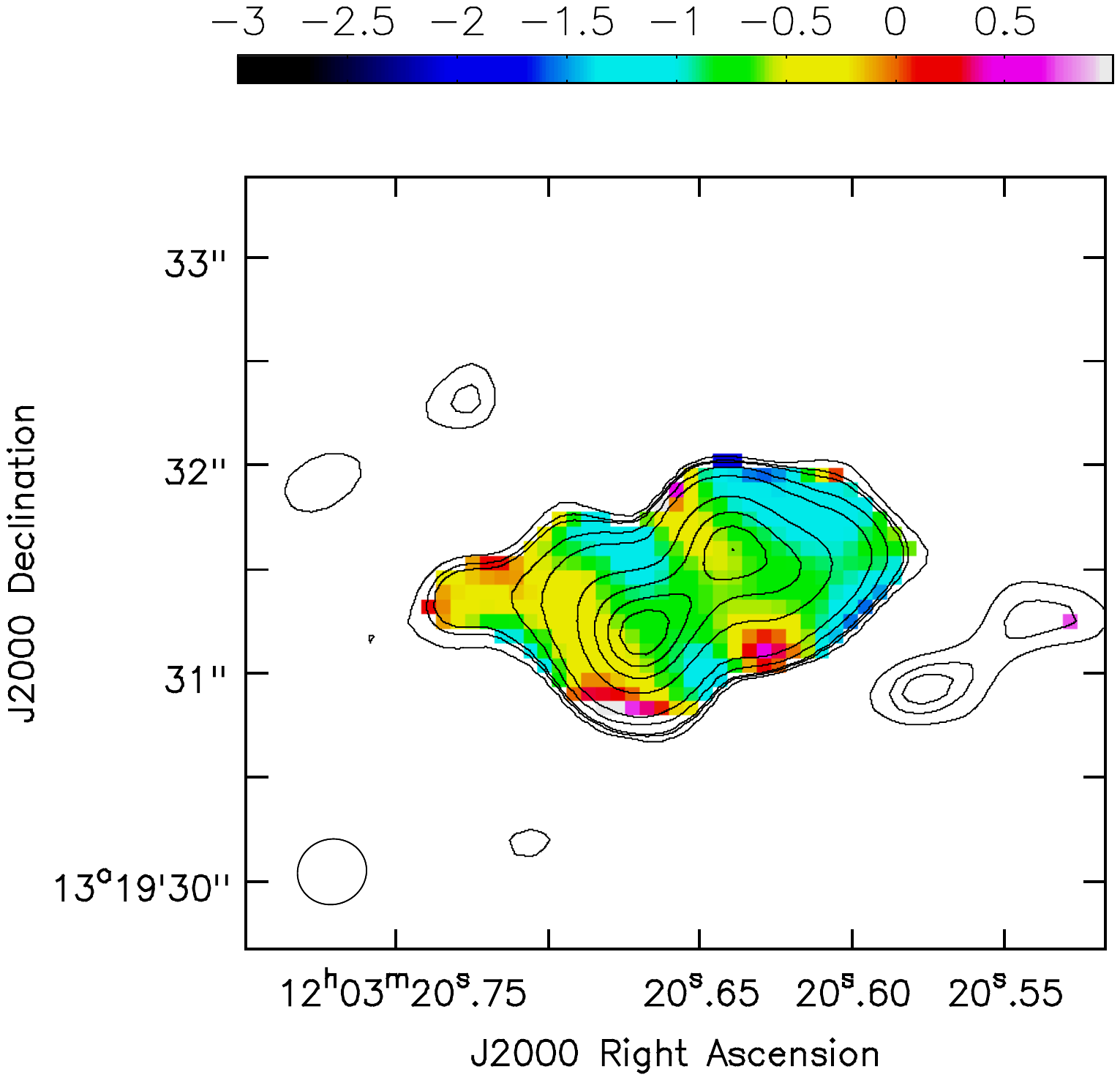}
\caption{\small The 6-15 GHz spectral index image in colour, 
superimposed by 15~GHz radio contours with levels corresponding to
60, 1.25, 2.5, 5, 10, 20, 40, 60, 80\%  of peak flux density value 
at 7.3~mJy. }
\label{fig9}
\end{figure}

The core has a spectral index value of $\alpha_{8.5}^{11.5}\sim-0.60\pm0.02$ between 8.5~GHz and 11.5~GHz 
which is moderately steep. Steep spectrum cores have sometimes been observed 
in Seyfert galaxies \citep{Peck.etal.2001}.
The SE hotspot has a spectral index of $\alpha_{8.5}^{11.5}\sim-0.76\pm0.01$ and
the NW one $\alpha_{8.5}^{11.5}\sim-0.72\pm0.02$ respectively 
(Figure~\ref{fig8}). We have obtained the spectral index values of jets using the 6~GHz and 15~GHz images. 
The values are $\alpha_{6}^{15}\sim-0.68\pm0.01$ and $\alpha_{6}^{15}\sim-0.61\pm0.01$ 
for the SE and NW hotspot respectively (Figure~\ref{fig9}).  
These spectral index values for the lobes are typical of lobes associated with jets in
large radio galaxies \citep[e.g.,][]{Laing.etal.1980,Dennett.etal.1999,Kharb.etal.2008}.

\subsection{Modeling the helical Jet-precession}\label{section5e}
\begin{figure}
\centering
\includegraphics[bb=0 0 1000 300,clip,scale=0.6]{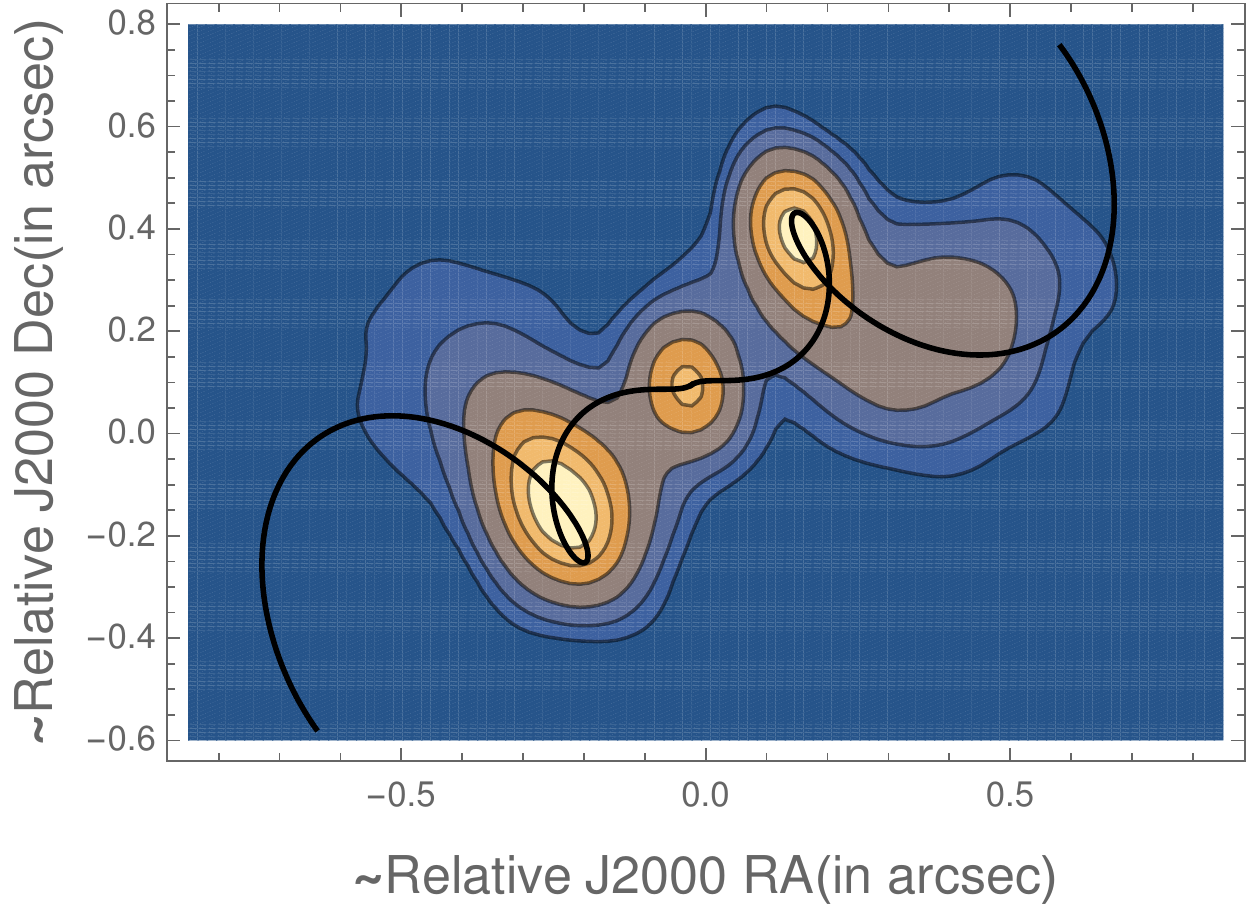}
\caption{\small The uniform weighted 11.5 GHz radio image of 2MASXJ12032061+131931 in color, superimposed 
by the precessing jet model of \citet{Hjellming.etal.1981} in black. 
The best-fit parameters are described in Section~\ref{section5e}.}
\label{fig10}
\end{figure}

S- or Z- shaped jets have been observed in many galaxies. \citet{Hutchings.etal.1988} found that almost 30~\% of all
quasars with z~$<$~1 and radio structures show an S-shape. Pronounced S-shaped jets are often termed helical jets and 
can be explained in terms of precessing jets \citep{Ekers.etal.1978,Parma.etal.1985}.
We have modeled the helical jets in 2MASXJ1203 using the \citet{Hjellming.etal.1981} model, since it is simple and does 
not assume any origin for the jet precession.
The model was used for SS433 but it can be used for extra-galactic jets as well.
From visual inspection we have derived the proper motion plot that matches the radio image in the following way.
We started with a typical jet advance speed of 0.03c \citep{Ulvestad.etal.2003} and varied the inclination angle {\it i} 
and half opening angle ($\psi$) from 0 to $90^{\circ}$. The best fitting model values are $i=52^{\circ}$, $\psi=21^{\circ}$, 
$v$~=~0.023c, precession period $P=0.95\times10^5$~years. Such precession timescales have been found 
for the jets of the Seyfert galaxy Mrk6 as well \citep{Kharb.etal.2006}.
The angle of rotation required to rotate the geometrical model to align it with the 11.5~GHz image is $\chi=33^{\circ}$. 
The proper motion is plotted for a jet precession timescale of $t=1.5\times10^5$~years  in Figure~\ref{fig10}.
To estimate the errors associated with our precession model 
we determined the range of parameters that can fit the radio morphology in the following way.
We fixed all the parameters except one and then varied this parameter until the
visually identified fit got significantly worse \citep{Steenbrugge2008}. The parameter ranges that 
we obtained were $i=52^{\circ}\pm5^{\circ}$, $\psi=21^{\circ}\pm2^{\circ}$ and $\chi=33^{\circ}\pm3^{\circ}$.
The range of precession period is  $P=(0.95\pm0.05)\times10^5$~years. The best fitting velocity which gives jets of the 
same size as in the radio image with the above parameters is $v=0.023c$. Small changes in jet velocity around 
this value do not change the shape much. But if $v$ is continuously increased, the jets start showing relativistic effects 
which make the jet structure look increasingly asymmetric.


One of the most common explanations for jet precession is the presence of a binary black hole system,
where the interaction of the black holes perturbs the spin of one or both black holes resulting 
in precessing jets \citep{Blundell.etal.2004,Romero.etal.2000,Roos.etal.1993}. Apart from binary black-hole systems, 
helical jets can arise when an accretion disk is irradiated by the central AGN and becomes 
unstable and warped thus producing jet precession \citep{Pringle.etal.1996,Livio.etal.1997};
it can also be due to the net change in the spin of the SBMHs which affects the orientation of the radio jets 
\citep{Rees.etal.1978,Natarajan.etal.1998}. Models with accretion disks of single AGN have also been
developed \citep{Lu.etal.1990}. Black-hole mergers can result in short-timescale 
redirection of the jet axis \citep{Merritt.etal.2005} and this may also lead to bent radio jets,
but these are X-shaped rather than S-shaped jets.

\subsection{Equipartition Estimates and Lifetimes}\label{section5f}
We have obtained the magnetic field corresponding to the minimum total energy of the synchrotron emitting plasma
which is close to the equipartition 
of the energy in particles and the magnetic field \citep{Burbidge1959}.
Using the radio luminosity \citep{Dea.1987}, we obtain the following expression.
\begin{equation}
L_{rad}=1.2\times10^{27}{D_L}^2S_0{\nu_0}^{-\alpha}(1+z)^{-(1+\alpha)}
({\nu_u}^{1+\alpha}-{\nu_l}^{1+\alpha})(1+\alpha)^{-1}\nonumber
\end{equation}
\begin{eqnarray} \nonumber
B_{min}=[2\pi(1+k)~c_{12}~L_{rad}~(V\phi)^{-1}]^{2/7}
\end{eqnarray}
Where L$_{rad}$ is the radio luminosity in erg~s$^{-1}$, D$_L$ is the luminosity distance in Mpc,
z is the redshift (Table~\ref{table1}), S$_0$ is the total flux density in Jy, $\nu_0$  is the frequency in Hz, 
$\nu_u$ and $\nu_l$ are the upper and lower cutoff frequencies respectively in  Hz,
k is the ratio of the relativistic proton to relativistic electron energy, V is the source volume, 
c$_{12}$ is a constant depending on the spectral index and frequency cutoffs \citep{Pacholczyk1970}, 
$\phi$ is the  volume filling factor 
and B$_{min}$ is the magnetic field at minimum pressure in Gauss. 
We have used the total flux density S$_0$~=~0.29~Jy from the 11.5~GHz~($\nu_0$) image (Figure~\ref{fig5}). 
The minimum magnetic field is calculated using the following assumptions: the radio spectrum extends 
from 100~MHz to 15~GHz with a spectral index value of $\alpha=-1$ (Figure~\ref{fig8}). 
The relativistic electron and proton have similar energy i.e k~=~1. The jets are uniformly filled with
relativistic particle and magnetic field i.e $\phi=1$. We have assumed cylindrical symmetry for the jet
and calculated the volume, $V~=~\pi(w/2)^2l$, using a length of $l\sim$~3~kpc and a width of $w\sim$~0.4~kpc.
The parameter $c_{12}$ is $6.5\times10^7$ for our observed frequency.
The estimated total radio luminosity is 1.9$\times$10$^{41}$~erg~s$^{-1}$.
We obtain a minimum magnetic field of 105~$\mu$G for a plasma filling factor of $\phi$~=~1.

We have used the relation from \citet{van1969} to estimate the lifetime of electrons in the radio component undergoing
both synchrotron radiative and inverse-Compton losses due to cosmic microwave background (CMB) photons,
\begin{eqnarray} \nonumber
t \approx \frac{2.6\times 10^4~{B_{min}}^{1/2}}{({B_{min}}^2+{B_{R}}^2)[(1+z)\nu]{^{1/2}}}
\end{eqnarray}
where B$_{min}$ is the minimum magnetic field from equipartition in G; B$_R\simeq~4\times10^{-6}(1+z)$~G is the
magnetic field equivalent to the radiation, which was assumed to be predominantly CMB photons and 
$\nu$ is the electron radiation frequency in Hz.
For the estimated magnetic field of 105~$\mu$G at 11.5~GHz we obtain the lifetime of electrons to be 
t~${\approx}$~2.1$\times$~10$^5$~yrs.

The age of the source can also be estimated via the spectral ageing analysis \citep{Myers1985}.
We have used Figure~3 from \citet{Myers1985} to obtain a rough estimate of X$_0$ which is 
the measure of maturity of synchrotron losses. We have used the $\alpha^{15}_{6}$ values (Figure~\ref{fig9}) at the initial hotspots 
and where jet falls below 3$\sigma$;
these frequencies match those used in Figure~3 of \citet{Myers1985}. We have used the Kardashav-Poacholczyk (KP) model 
where assumes an isotropic pitch angle distribution. The initial electron energy is from synchrotron radiation
which follows a power law.
The magnetic field is assumed to be the equipartition value.  
The estimated timescale is t~${>}$~1.3$\times$~10$^5$~yrs. 
This is a lower limit since only the hotspot and the tail were considered instead of the entire core-jet structure.
We have calculated the jet speed using the synchrotron time i.e ~2$\times10^5$~yrs 
and the size of one-side jet i.e ~1.5~kpc which comes out to be $\beta$~=~0.023. 
The calculated jet advance speed is the same as was obtained from the precession model in Section~\ref{section5e}. 
The timescale of the order of $\sim10^5$ is also consistent with the precession timescale obtained in Section~\ref{section5e}.

\section{DISCUSSION} \label{section6}

Our radio observations show that 2MASXJ1203 has a  core-jet structure with S-shaped or helical jets.
In this section we discuss the mechanisms that could explain the radio observations of the DPAGN in 2MASXJ1203.
 
\subsection{Mechanisms Causing S-shaped Core Jet Structures} \label{section6a}
In the literature, Z or S-shaped radio galaxies are referred to as X-shaped galaxies \citep{Cheung2007}.
However, it is very difficult to explain this morphology with the models for X-shaped galaxies which consist
mainly of (i)~spin flip of a SMBH \citep{Natarajan.etal.1998} or (ii)~backflow of plasma in radio jets \citep{Leahy.etal.1984}.
In the back-flow model, the ISM of the host elliptical galaxy exerts a buoyancy pressure 
on the back-flowing plasma causing secondary wings of plasma to form from the terminal shocks
evolving in the hot medium \citep{Capetti2002}. The rapid cocoon expansion along the direction of the hostâs
minor axis produces the X-shaped morphology. The relationship between  the optical axis  and  wing  orientation indicates 
that the formation of the XRSs is intimately related to the host galaxy's geometry \citep{Gillone2016}. 
These results strengthen the interpretation that the X-shaped morphology in radio-sources has a hydrodynamical origin. 
The back-flow model can explain the radio structures associated with the large FR I \citet{Fanaroff1974} galaxies.
But the model is not able to explain why the secondary lobes are larger than the primary lobes in many radio galaxies
such as NGC 326, 4C+00.58 \citep{Hodges2010}. However, 2MASXJ1203 cannot be classified as 
an FRI radio source due to its small size and the secondary wings are larger than the primary.
There is also no evidence of cocoon formation in the radio images. 
Therefore, the back-flow model cannot explain the radio morphology of 2MASXJ1203. \citet{Gopal.etal.2003} have
proposed a different model which can explain both X and Z-shaped sources. In this model, as a galaxy spirals
into the nucleus of the host galaxy, it perturbs the axisymmetric disk potential, causing disk rotation in the 
interstellar medium (ISM) of the host galaxy \citep{Noel.etal.2003}. The gas in the disk will interact with the jets
causing them to bend at large radii - this may appear as jet precession i.e. Z or X shaped radio jets. 
A good example is the Z-shaped galaxy NGC 3801 \citep{Das2005}. \citet{Hota.etal.2009} showed that Z-shaped radio jets 
in NGC~3801 have gone through a recent merger and it has a large fast rotating gas disk interacting with jets.
In our case, however, there is no indication of jet-ISM interaction and our radio jets have small radii of $\sim 1.5$~kpc. 
Thus, this model is probably not the explanation for the S-shaped radio jet morphology in 2MASXJ1203.

A helical jet structure can also be due to jet precession during the lifetime of a radio source \citep{Ekers.etal.1978}.
The jet precession model fits the radio image of 2MASXJ1203 (see Section~\ref{section5e}) and the calculated jet
precession timescale is similar to the age of the electrons in the lobes i.e $10^5$~yrs (see Section~\ref{section5f}).
Therefore,  jet precession is the most probable explanation for the S-shaped radio structure of 2MASXJ1203.
We discuss jet precession in greater detail in the next section.

\subsection{Mechanisms Causing Jet Precession} \label{section6b}
S-symmetry due to precession has long been predicted associated with the presence of binary SMBHs \citep{begelman.etal.1980}.
The secondary BH can induce the rapid precession in the inner region of the primary accretion disk \citep{Romero.etal.2000}.
The motion of the secondary BH around the primary BH can also cause the orbit to precess resulting in helical or S-shaped radio
jets \citep{Roos.etal.1993,Valtonen2016}. A good example is the BL Lac object Mrk 501 that has helical jets 
which can be explained by binary SMBH models \citep {Villata.etal.1999}. 
We have calculated the separation of the binary system which can give the precession period of $10^5$~years. 
The relation, $P_{prec}\sim600~{r_{16}}^{5/2}(M/m)~{M_8}^{-3/2}$~yr \citep{begelman.etal.1980}
gives a separation of $\sim0.02$~pc, where $r_{16}$ is the separation in units of $10^{16}$~cm,
$M_{8}$ is the mass of the primary SMBH in units of $10^{8}~M_{\odot}$ and (M/m) is the mass ratio of primary SMBH 
to secondary SMBH. We have used the mass ratio 1.084 and $M_8\sim0.82$  (Section~\ref{section5c}). 
Therefore, there could be a second SMBH  at a separation of 0.02 pc in 2MASXJ1203, 
which we cannot detect with the present day telescopes.  
The other possibility could be that there exists a dual SMBH system and the secondary SMBH
has passed the primary SMBH in the past; this could have induced the precession in the jet of the primary SMBH. 
Here the secondary SMBH may not have sufficient radio flux density to be detected in our observation; 
higher resolution observations however, could detect it, if present. In the case of 2MASXJ1203, 
there appears to be a second NIR Core in  Keck NIRC2 image \citep{McGurk.etal.2015}. 
The image contains a second core at a separation of 2.1~kpc from the nucleus.
However, spectroscopic observations have not been carried out. 
The second faint core could form a DAGN system with the nuclear SMBH in the bulge.
However, we cannot rule out an AGN-starburst system either. If the faint NIR core is a SMBH it does not have enough
radio flux density to be detected in our maps. This second SMBH could cause the jet precession in 2MASXJ1203.

Though binary models are interesting, a single AGN model can also explain the helical jet morphology.
On the basis of  the \citet{Sarazin.etal.1980}  model for SS433, \citet{Lu.etal.1990} 
suggested that a tilted accretion disk can also produce jet precession. This model of a tilted accretion disk
around a single AGN gives a relation between the precession period and the luminosity.
The radio source 1946+708 has jets with a tilted disk \citep{Peck.etal.2001} and the observed morphology follows
the period-luminosity relation of \citet{Lu.etal.1990}.
The precession period is usually calculated using several precession models which have very large
uncertainties and the magnitude values also differ from literature to literature.
We have used the  period-luminosity relation where 
the absolute B-band magnitude of our target\footnote{http://leda.univ-lyon1.fr/} is $M_{abs}$~=~-20.34 and
the calculated precession period is ${\sim}10^{5}-10^9$~yrs.
We note that the timescale uncertainties are very large. However, our jet precession model time estimate of ${\sim}10^{5}$~yrs 
that was obtained in Section~\ref{section5e}, falls within this time range. It is possible that the accretion disk in 
2MASXJ1203 has become warped or tilted due to non-uniform irradiation from the AGN. 
Such radiation can cause the accretion disk to become unstable and warped, 
resulting in jet precession \citep{Pringle.etal.1997}. 
In radio images we have not detected the second core so we cannot distinguish between these two mechanisms of precession. 
Thus in summary the jet precession in 2MASXJ1203 can be due to a close SMBH binary at a separation of ~0.02~pc,
a dual system in which a close passing of an SMBH has induced the precession or a single AGN with 
an warped accretion disk that has given rise to jet precession.

\subsection{Is 2MASXJ1203 a CSS/CSO source~?} \label{section6c}
The other class of radio sources into which 2MASXJ1203 falls is that of the compact steep spectrum (CSS) objects
or compact symmetric objects (CSO). Gigahertz peaked-spectrum (GPS)($\leq$1 kpc) and CSS ($\leq$20 kpc) sources 
are small bright sources with steep spectra and are young ($\leq 10^5$ yrs). 
CSS and GPS sources can probe the NLR of the host galaxy \citep{Dea1998}. 
CSO are radio objects which have compact symmetric double lobes that extend to small
galactic radii of $\leq$ 1~kpc \citep{Wilkinson.etal.1994}. CSO are a subclass of GPS with symmetric jets.
CSS are also symmetric in structure. \citet{Fanti1995,Readhead1996} have proposed that GPS and CSS are
the evolutionary stages of large radio galaxies i.e GPS $\rightarrow$ CSS $\rightarrow$ large radio galaxies. 
Some CSOs show S-symmetry \citep{Deane.etal.2014} which can be signature of binary SMBHs \citep{begelman.etal.1980}.
The binary AGN with the smallest separation, 0402+379, is a CSO \citep{Rodriguez.etal.2006}. \citet{gopal1995} have
suggested that a merger, either with a dwarf galaxy or a secondary SMBH, could disturb the torus funnel in the AGN 
and give rise to jet misalignment. Such a distorted torus can give rise to a GPS source.
We have calculated the break frequency of 2MASXJ1203 using $t_s=1610B^{-3/2}\nu_b^{-1/2}$~Myr
from \citep{Carilli1991}, where t$_s$ is the synchrotron age , B is the magnetic field and $\nu_B$ is 
the break frequency. We have used  t$_s$= 10$^5$ yrs, B$=105\mu$G from Section~\ref{section5f}. 
The calculated break frequency is $\nu_B\sim0.55$~GHz. This $\nu_B$ is in the break frequency range observed 
in CSO/CSS sources \citep{Murgia2003}. The radio observations of 2MASXJ1203 show an S-shaped jet of
total extent $\sim$~3~kpc, the average spectral index $\alpha_{8.5}^{11.5}\sim-1$,~ 
 $\nu_B$ is 0.55~GHz and the estimated age is $\sim10^5$~yrs. Thus 2MASXJ1203 is a 
CSS/CSO source and its S-symmetry could be consistent with the binary/dual SMBH scenario.

\subsection{Origin of Double-peaked [O~III] line} \label{section6d}
Double peaked emission lines from the NLR have been detected in nuclear galaxy spectra 
since the 1970s \citep{Sargent.etal.1972,Heckman.etal.1981} and are important for studies of NLR kinematics.
It was previously thought that DPAGN emission lines are indicators of bipolar outflows or
rotating disks \citep{Greene.etal.2005}. However, they are now thought to also indicate the presence of
binary or dual AGN \citep{Zhou.etal.2004} at the separations of $\sim$100~pc to $\sim$10~kpc 
respectively \citep{Wang.etal.2009}. Since the data of large spectroscopic surveys such as SDSS 
became available \citep{Smith.etal.2010, Fu.etal.2012}, there has been a greater interest in finding large
sample of DPAGNs since they may represent DAGN. However, there are several other mechanisms that can 
produce double-peaked emission lines from a single AGN \citep{Xu.etal2009}. For example, jet-cloud interaction
can also produce a DPAGN \citep{Rosario.etal.2010}. \citet{Smith.etal.2010} suggest that if two Gaussian 
components of a double-peaked [O~III] line are symmetric than it can be explained by a rotating ring. Thus,
it is very difficult to confirm the presence of dual/binary AGN just from DPAGN emission and one has to carry
out high resolution imaging to confirm it. \citet{Tingay.etal.2011} conducted a VLBI study of DPAGN to
search for dual/binary AGN but they have not detected any. \citet{Comerford.etal.2012} used an additional 
criteria to detect DAGN from a sample of DPAGN; they used
long-slit spectroscopy to separate the outflow and rotating disks from candidate dual/binary AGN sources.
They found 17 promising dual AGN candidates from 81 DPAGN sources. 
\citet{Mullersanchez.etal.2015} have tried to constrain the origin of DPAGN using VLA observations along 
with long-slit spectroscopic studies. 
They find  that 15\% of their sample are dual AGN and in 75\% the DPAGN are due to gas kinematics.
From these recent studies, it appears that DPAGN may not be promising candidates for dual/binary AGN. 

In our study, we do not have a two dimensional [O~III] map of the gas kinematics or high resolution long slit observations for 
our target. Therefore, we cannot be sure about the origin of the double-peaked emission lines in 2MASXJ1203 just from the SDSS spectrum.
However, just to check if the double peaked [O~III] lines could be due to two SMBHs, we used the velocity separation 
$\Delta v~=~292~km~s^{-1}$ of the [O~III] emission lines \citep{Ge.etal.2012} to make an approximate estimate of the SMBH separation using 
$\Delta R~\sim~GM/(\Delta v)^{2}$ and used it to derive a precession timescale. The calculated separation for 
$M=10^{8}~M_{\odot}$ is ~$\sim8$~pc. Using this value in equation 7 from \citet{begelman.etal.1980}
gives a precession period of ~$4.6\times10^{11}$~yrs which is more than the Hubble time. 
This value contradicts the model precession timescale (which is due to a binary of separation of 0.02~pc) 
as well as the synchrotron lifetimes 
of electrons in the radio jets, both of which are of the order of $\sim10^5$~yr.
Thus in the precession model it is not possible that  
a close binary is the origin of the double peaked emission lines in the SDSS spectrum of 2MASXJ1203. 
However, it could be that there is a dual system in which the precession has been induced during a close pass;
in that case we cannot rule 
out the possibility that the DPAGN is due to two AGN. Alternatively, there is only a single AGN in which case
the jet-ISM interaction is the 
probable origin of the DPAGN emission line. The kpc radio jets may be responsible for
the double-peaks through jet-medium interaction.

\subsection{Detecting the dual/binary AGN in 2MASXJ1203} \label{section6e}

The S-shaped radio morphology of  2MASXJ1203 is due to slow precession of jets during its lifetime and it is 
 a CSS/CSO source. We have considered three scenarios for jet precession: (i) binary AGN at the separation of 0.02~pc, 
 (ii) a dual AGN system where a close pass of the secondary SMBH in the past has given rise to the jet precssion or
 (iii) a single AGN with a tilted accretion disk. If the first case is true, 
 i.e., the separation is 18 microarcseconds in the sky, 
 we cannot resolve the second AGN with current ground-based VLBI telescopes. 
 With our present data we are not able to
 rule out any of these possibilities but future higher resolution observations may help us if 
 the SMBHs are at separation lying between $40\leq$d$\leq100$~pc. If the separation is more than 100~pc,
 the second AGN does not have sufficient radio flux density since it has not been detected in our observations.

\subsection{The relevance of DPAGN to dual/binary AGN studies} \label{section6f}
Recent high resolution radio observations have shown that DPAGN are not 
usually dual/binary AGN \citep{Shen2011,Fu2011b} as indicated by earlier studies and DAGN appear to be far 
more elusive than thought. However, they are very important for understanding the nuclear disk kinematics and 
good indicators of kpc scale radio jets or outflows \citep{muller.etal.2011,Kharb.etal.2015}.
The lack of parsec scale resolution and the absence of radio or X-ray emission in a significant fraction of
AGN makes it harder to detect DAGN. In some cases the NIR images or radio morphologies may have indirect signatures 
of the SMBH pair - such as lopsided disks or S-shaped jets as in 2MASXJ1203. 
These cases need to be followed with higher resolution observations. Thus multi-wavelength
and high resolution observations are essential to detect dual/binary AGN in DPAGN 
galaxies.


\section{Summary} \label{section7}
We summarise the conclusions of our study below:\\
{\bf 1.~} We have carried out radio imaging of the double-peaked emission line Seyfert~2 galaxy 
2MASXJ1203 using EVLA at 6, 8.5, 11.5 and 15 GHz. 
The 6 and 15~GHz images show two distinct radio hotspots on either side of the optical nucleus. \\
{\bf 2.~}The 8.5 and 11.5 GHz images obtained via archival data,
reveal the full extent of the emission and we are able to resolve the core jet structure. 
The radio jets have an S-shaped helical structure extending out to a radius of 
$\sim1.5^{\prime\prime}$ (1.74~kpc) on either side of a deconvolved core of size $\sim0.1^{\prime\prime}$ (116~parsec). 
   \\ 
{\bf 3.~}We have modeled the helical-jet structure using the \citet{Hjellming.etal.1981} model. 
The best-fit jet advance speed is 0.023c and precession timescale is $\sim10^5$~yrs. 
The half opening angle is $\psi=21^{\circ}\pm2^{\circ}$ and inclination angle 
for the radio jets is  $i=52^{\circ}\pm5^{\circ}$.
\\
{\bf 4.~} We have calculated the minimum magnetic field value of 105~$\mu$G and
the electron lifetime of $\sim10^5$ years from the equipartition theorem.
This timescale matches the time the precessing jet was "on" in this Seyfert galaxy, providing support to the precession model.\\
{\bf 5.~} 2MASXJ1203 is compact steep spectrum/compact symmetric object. 
Such sources have been suggested to be binary AGN in the literature.\\
{\bf 6.~} The presence of S-shaped precessing radio jets in 2MASXJ1203 can be due to binary/dual SMBH or 
a single tilted SMBH with accretion disk.
Double-peaked emission lines also can be due to binary/dual AGN or NLR kinematics of a single AGN. 
While the binary/dual SMBH scenario is supported by several suggestions, 
we are unable to rule out other possibilities with the present data.
 Future high resolution multi-wavelength (radio, X-ray, optical) observations are required to 
get a clearer picture for the double-peaked AGN, 2MASXJ1203.\\

 We thank the referee for insightful comments that improved the paper.
We  acknowledge IIA for providing the computational facilities.
Rubinur K. wants to thank Dr. Avijeet Prasad for helpful discussions.
The National Radio Astronomy Observatory is a facility
of the National Science Foundation operated under cooperative
agreement by Associated Universities, Inc.
This research has made use of the NASA/IPAC Extragalactic Database (NED),
which is operated by the Jet Propulsion Laboratory, California
Institute of Technology, under contract with the National Aeronautics and Space Administration.
Funding for the Sloan Digital Sky Survey IV has been provided by 
the Alfred P. Sloan Foundation, the U.S. Department of Energy Office of Science,
and the Participating Institutions. SDSS- IV acknowledges support and resources 
from the Center for High-Performance Computing at the University of Utah. 
The SDSS web site is www.sdss.org.
We acknowledge the usage of the HyperLeda database (http://leda.univ-lyon1.fr)


\bibliographystyle{mn2e}
\bibliography{ms}

\begin{table*}
\centering
\caption{ Sample Galaxy}
\vspace{0.5cm}
\label{table1}
\begin{tabular}{|l|l|l|}
\hline
Parameters         & Values                        &   Reference      \\\hline
Galaxy name        & 2MASX J12032061+1319316       &   NED         \\ \hline
RA                 & 12:03:20.7658                    &   NED         \\ \hline
DEC                & +13:19:31.39                    &   NED         \\ \hline
Galaxy Type        & S0                            &   \citep{Fathi.etal.2010}. HyperLeda         \\ \hline
Redshift           & 0.058423                          &   NED         \\ \hline
Luminosity Distance & 245.4~$\pm$~17.2 Mpc            &   NED        \\ \hline
Radio Loudness Parameter     & 156                           &   \citep{Fu.etal.2012}         \\ \hline
1.4 GHz Peak Flux Density      & 0.103~Jy/beam                      &   NVSS (beam~=~45$^{\arcsec}$ )       \\ \hline
Optical Absolute Mag (z band)     & -22.07~$\pm$~ 0.50          &   SDSS  {DR9}       \\ \hline
Velocity Dispersion      &  189.9~km~s$^{-1}$                &  pPxf using SDSS DR9 spectra        \\ \hline
\end{tabular}
\end {table*}

\begin{table*}
\centering
\caption{Observation Details }
\vspace{0.5cm}
\label{table2}
\begin{tabular}{|l|l|l|l|}
\hline
Frequency & Observing Date             & ID          & Reference \\ \hline
6~GHz     & 20 July, 2015   & VLA/15A-068 & Our Data  \\ \hline
8.5~GHz   & 14 March, 2014  & VLA/13B-020 & Archival Data   \\ \hline
11.5~GHz  & 14 March, 2014 & VLA/13B-020 & Archival Data  \\\hline
15~GHz    & 29 May, 2016   & VLA/16A-144 & Our Data \\ \hline
\end{tabular}
\end{table*}

\onecolumn
\begin{table}
\centering
\caption{Radio Properties of 2MASXJ12032061+131931}
\vspace{0.8cm}
\label{my-label}
\resizebox{1.0\textwidth}{!}{
\begin{tabular}{|c|c|c|c|c|c|c|c|c|c|}
\hline
&\begin{tabular}[c]{@{}l@{}}Array \\configuration \end{tabular}
&\begin{tabular}[c]{@{}l@{}}Robust\\parameter \end{tabular} 
& \begin{tabular}[c]{@{}l@{}} Image  \\ Noise \\in $\mu$Jy \end{tabular}
& \begin{tabular}[c]{@{}l@{}}Core Size in \\ $\theta_M\times\theta_N$~($\prime\prime$)\end{tabular}
&\begin{tabular}[c]{@{}l@{}}Core Peak \\ Flux in mJy\end{tabular}
&\begin{tabular}[c]{@{}l@{}} East Jet \\ size in ${\prime\prime}$ \end{tabular} 
&\begin{tabular}[c]{@{}l@{}} South East  \\Hotspot  Peak \\Flux in  mJy \end{tabular}  
& \begin{tabular}[c]{@{}l@{}}West Jet \\size in ${\prime\prime}$ \end{tabular} 
& \begin{tabular}[c]{@{}l@{}}North West \\Hotspot Peak  \\Flux in mJy\end{tabular} \\\hline
6~GHz            & A              &  -0.5    & 21     & not resolved & --    & 1.50   & 11.90   & 1.03    & 8.40      \\ \hline
8.5~GHz          & A              &  -0.5    & 14     & 24$\times$18 & 3.8   & 1.11   & 6.86   & 1.20    & 5.68      \\ \hline
8.5~GHz          & A              &   0.5    & 17     & Not resolved & --    & 1.50   & 8.20   & 1.45   &   6.40    \\ \hline
11.5~GHz         & A              &  -0.5    & 19     & 20$\times$14 & 3.2   & 1.05   &  4.90  &  0.96   &  4.44     \\ \hline
15~GHz           & B                 &  -0.5    & 13     & Not resolved  & --   & 0.90   & 7.30 & 0.90 & 5.00 \\ \hline

\end{tabular}
}
\label{table3}
\end{table}

%

%
%
%
%
%
%
%



\end{document}